\documentclass[epsf]{elsart}

\usepackage[dvips]{graphicx}
\graphicspath{{./}}

\begin{document}
\begin{frontmatter}
\title{Effect of large neutron excess on the 
dipole response \\
 in the region of the Giant Dipole 
Resonance }

\author[Catania,Sevilla]{F. Catara}, 
\author[Catania]{E.G. Lanza}, 
\author[Legnaro]{M.A. Nagarajan\thanksref{man}} and 
\author[Padova] {A. Vitturi}
\address[Catania]{ Dipartimento di Fisica and INFN, Catania, Italy}
\address[Legnaro]{Laboratori Nazionali di Legnaro, INFN, Italy }
\address[Padova]{ Dipartimento di Fisica and INFN, Padova, Italy}
\address[Sevilla]{Departamento de Fisica Atomica, 
Molecular y Nuclear, Sevilla, Spain}
\thanks[man]{Current address: Department of
Physics, UMIST, Manchester M60 IQD, UK} 

\begin{abstract}
The evolution of the Dipole Response in nuclei with strong neutron
excess is studied in the Hartree-Fock plus Random Phase Approximation
with Skyrme forces. We find that the neutron excess increases the
fragmentation of the isovector Giant Dipole Resonance, while pushing
the centroid of the distribution to lower energies beyond the mass
dependence predicted by the collective models. The radial separation
of proton and neutron densities associated with a large neutron excess
leads to non vanishing isoscalar transition densities to the GDR
states, which are therefore predicted to be excited also by isoscalar
nuclear probes.  The evolution of the isoscalar compression dipole
mode as a function of the neutron excess is finally studied.  We find
that the large neutron excess leads to a strong concentration of the
strength associated with the isoscalar dipole operator
$\sum_ir^3_iY_{10}$, that mainly originates from uncorrelated
excitations of the neutrons of the skin. 
\end{abstract}
\end{frontmatter}

Currently, there has been interest in the study of the effect of
neutron skin on collective states in neutron-rich nuclei~\cite{1,2,3,4}.
One of the collective modes of interest is the Isovector Giant Dipole
resonance (GDR) in such nuclei.  In neutron-rich nuclei, one expects
the neutron and proton densities to have different shapes.  This would
allow the possibility of exciting the isovector GDR by isoscalar
probes through hadronic interactions~\cite{5,6,7}, in addition to the
usual method through Coulomb excitation~\cite{8}.  The excitation of
the isovector GDR by isoscalar probes through hadronic interactions is
possibly the most sensitive measure of the ``neutron skin''.  This
property can be assessed by the evaluation of the isoscalar and
isovector dipole transition densities in neutron-rich nuclei.

In order to investigate the effect of neutron skin on collective
dipole states of neutron-rich nuclei, we have performed microscopic
calculations for several isotopes of oxygen and calcium nuclei, based
on spherical Hartree-Fock (HF) method with Skyrme SGII interaction.
The proton and neutron densities in $^{28}$O and $^{60}$Ca are shown
in Fig.~1.  The HF calculations predict the last neutron to be bound
by 3.25 MeV in $^{28}$O and 5.1 MeV in $^{60}$Ca.  These relatively
large values of the neutron separation energy hinder the occurrence of
``unusual'' concentration of dipole strength at the continuum
threshold, an effect that is directly associated with very small
binding energy~\cite{3,9}.  Note that the use of other Skyrme
interactions may give rather smaller neutron separation energy,
leading to different features in the very low energy part of the
response.  This should not alter the medium and high energy regions,
which are the object of the present investigation.

\begin{figure}
\begin{center}
\includegraphics[bb= 110 110 550 719,angle=-90,scale=0.6]{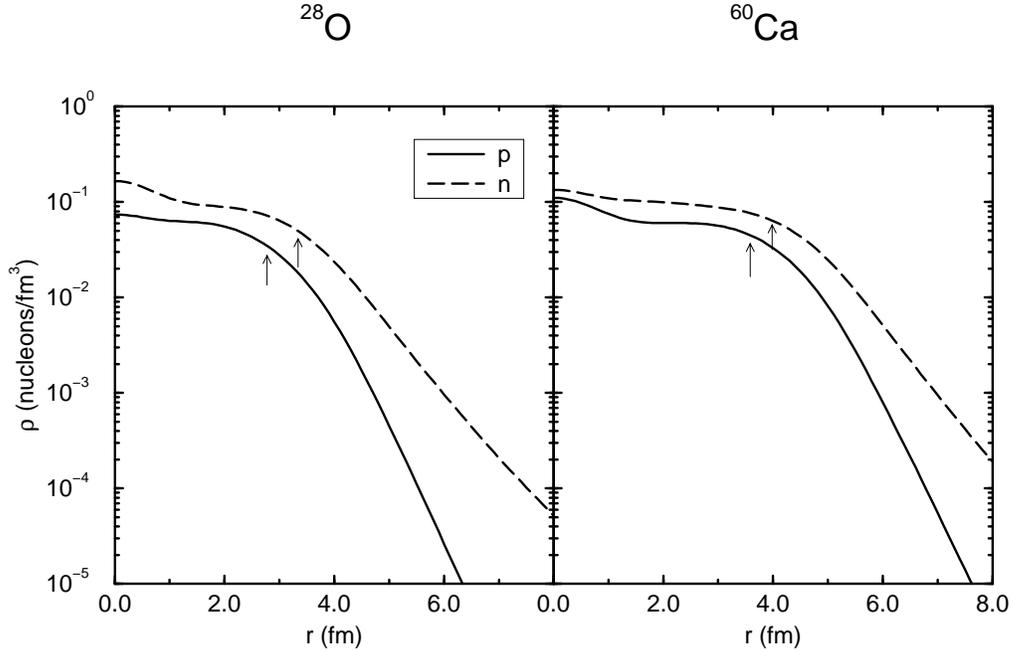}
\end{center}
\caption 
{ Hartree-Fock proton and neutron 
densities obtained for
$^{28}$O and $^{60}$Ca with SGII interaction. The arrows indicate the
r.m.s. radii.}
\end{figure}

The collective dipole excitations of these nuclei were calculated in
RPA, using the full residual interaction.  After obtaining the dipole
states by the RPA calculation, we calculated the response to the
isovector dipole operator $\sum_i z_i\tau_i^3$.  Similarly we can
calculate the isoscalar response to the operator $\sum_i z_i$, which
should vanish if the RPA states were proper eigenstates of a
translationally invariant hamiltonian, together with the occurrence at
zero excitation energy of the spurious state.  Since the last neutrons
in these nuclei are not too weakly bound, the continuum states needed
for the RPA were expanded in oscillator functions of different
principal quantum number, a procedure which was found to be adequate
in these systems.  In actual RPA calculations, because of the
truncation involved and the additional effect of the isospin mixing
introduced by the Coulomb interaction, the spurious isoscalar mode
occurs at very low excitation energy (carrying a large fraction of its
strength) and there are insignificant spurious center-of-mass
components at higher energies.  The spurious state can thus be
eliminated.

\begin{figure}
\begin{center}
\includegraphics[bb= 90 90 550 719,angle=-90,scale=0.7]{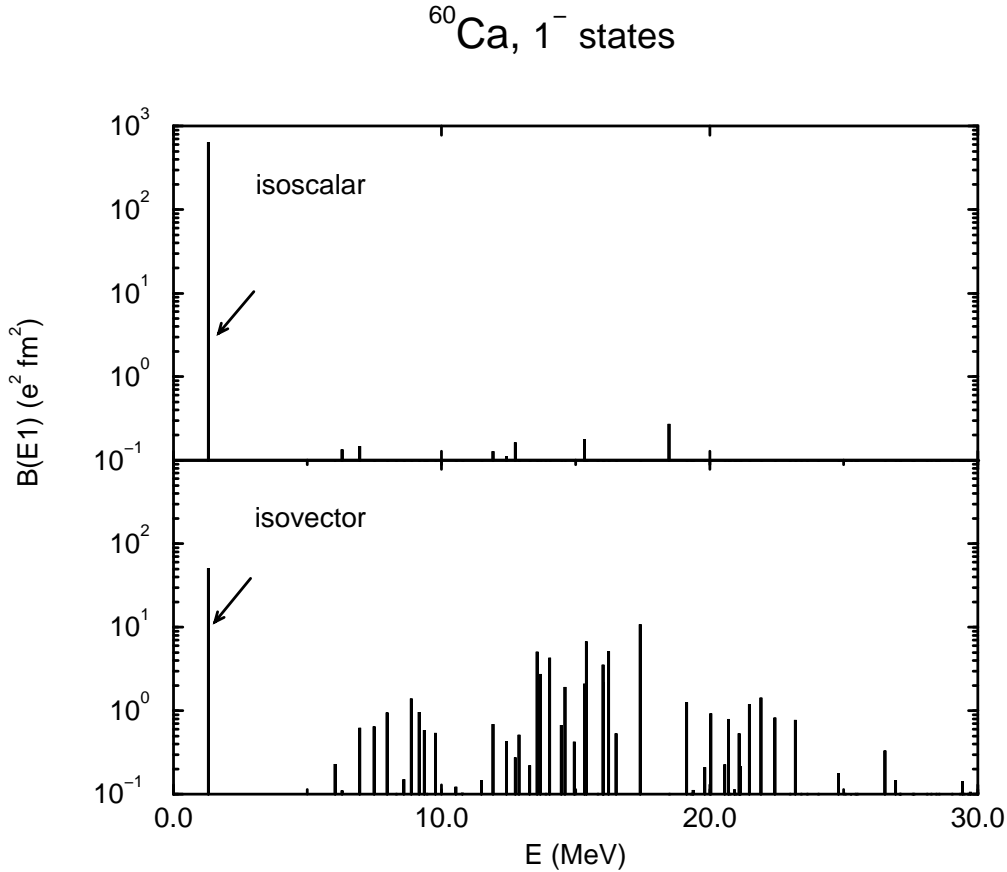}
\end{center}
\caption 
{ Isoscalar and isovector dipole response 
obtained in HF+RPA
for $^{60}$Ca. Most of the spurious isoscalar response 
is concentrated in
one low-lying state which can be easily eliminated. }

\end{figure}

An example of the calculated $B(E1)$'s for isoscalar and isovector
dipole strength distributions is shown for $^{60}$Ca in Fig.~2.  The
spurious state occurring at low excitation energy is marked with an
arrow.

\begin{figure}
\begin{center}
\includegraphics[bb= 80 80 550 719,angle=-90,scale=0.6]{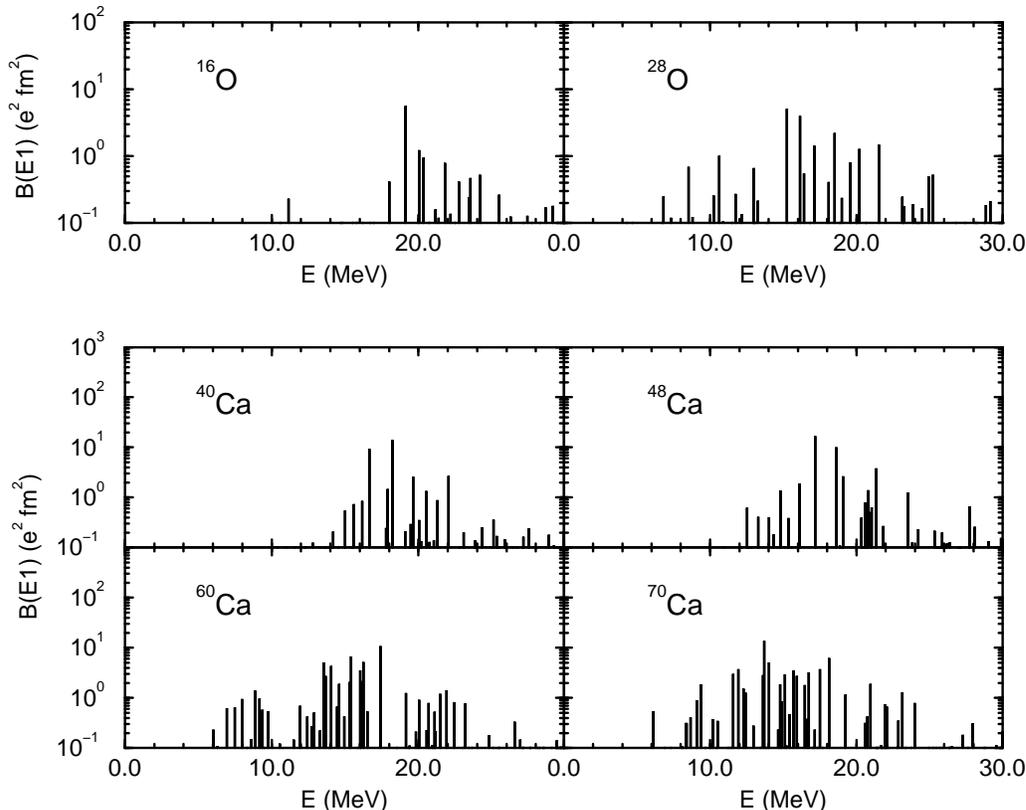}
\end{center}
\caption 
{Isovector dipole response obtained in HF+RPA for a sequence of Oxygen
and Calcium isotopes.  Spurious states have been eliminated
(cf. Fig.~2). }

\end{figure}

In view of the interest in the effect of neutron excess on the
isovector dipole states, the RPA results of the different oxygen and
calcium isotopes are shown in Fig.~3.  One of the effects of the
neutron excess is the spreading of the isovector dipole strength.  It
should be noted that in both the $^{28}$O and $^{60}$Ca cases, the
most ``collective'' state only exhausts approximately 15 \% of the
total EWSR.  The increased spreading in the neutron rich isotopes is
more clearly seen in Fig.~4, where we have averaged the dipole
response with a lorentzian with a width of 2 MeV.  In both cases the
full width at half maximum (FWHM) is seen to have increased by about
50 \% going from the $N$=$Z$ to the most neutron rich isotope.  In
addition, one observes that the centroid of the strength function
shifts to lower energy in the neutron rich isotopes.  This shift is
larger than can be accounted by the $A^{-1/6}$ dependence given by the
Goldhaber-Teller~\cite{10} model. For example the centroid changes from
19 MeV in $^{40}$Ca to 16 MeV in $^{60}$Ca, a shift which is a factor
two larger than the prediction of the Goldhaber-Teller model.  The
shift is closer, although still slightly larger, to the prediction of
the $A^{-1/3}$ scaling associated with the hydrodynamical
model~\cite{11}.

\begin{figure}
\begin{center}
\includegraphics[bb= 90 90 550 719,angle=-90,scale=0.6]{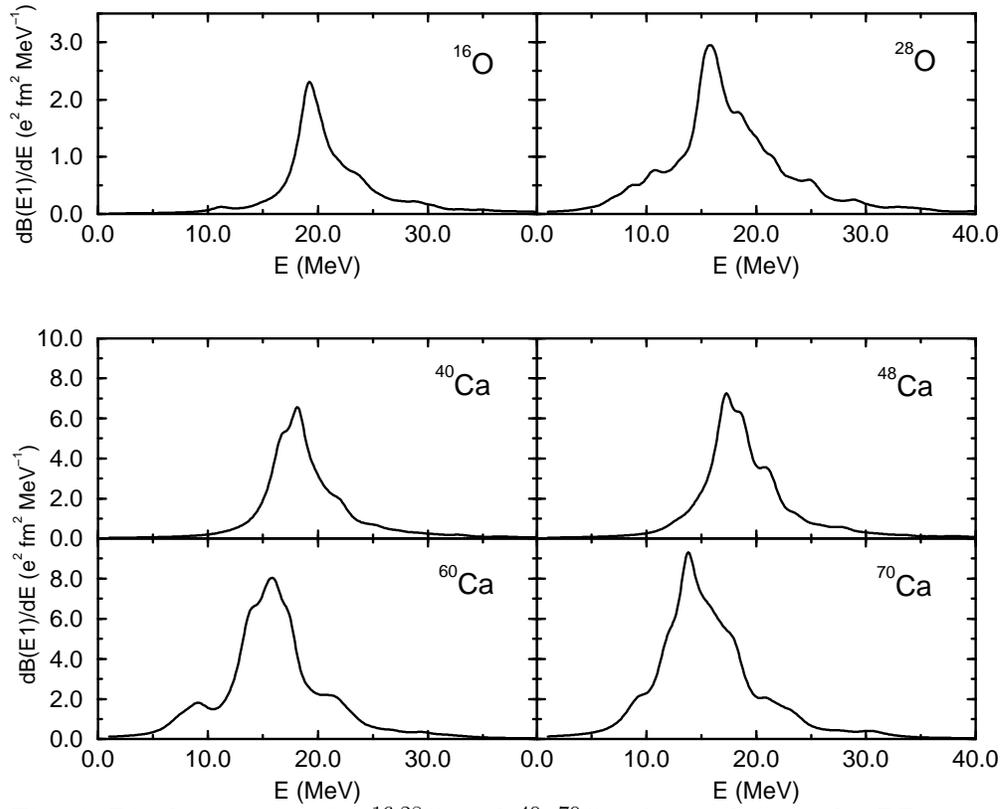}
\end{center}
\caption 
{Dipole response for $^{16,28}$O and $^{40-70}$Ca obtained from the
RPA response shown in Fig.~3, by averaging the discrete spectra with a
lorentzian with $\Gamma$=2 MeV. }

\end{figure}

Even though the isoscalar $B(E1)$ to all states must identically
vanish, the corresponding isoscalar transition densities to the
different states need not to be identically zero. For example, within
the collective Goldhaber-Teller model for the GDR, only in the
particular case where the neutron and proton densities have the same
shape (and scale with $N$ and $Z$) would the isoscalar dipole
transition density vanish.  If the neutron and proton transition
densities have different shapes, as is the case for very neutron rich
nuclei, the corresponding isoscalar dipole transition density will be
non zero.  The isoscalar and isovector-dipole transition densities to
a selected state in $^{28}$O and $^{60}$Ca are shown in Fig.~5.  The
states selected have 18 and 17.4 MeV of excitation energy and carry 8
and 16 \% of the EWSR, respectively.  In these figures also are shown
the separate neutron and proton transition densities.  

\begin{figure}
\begin{center}
\includegraphics[bb= 10 10 550 719,angle=0,scale=0.6]{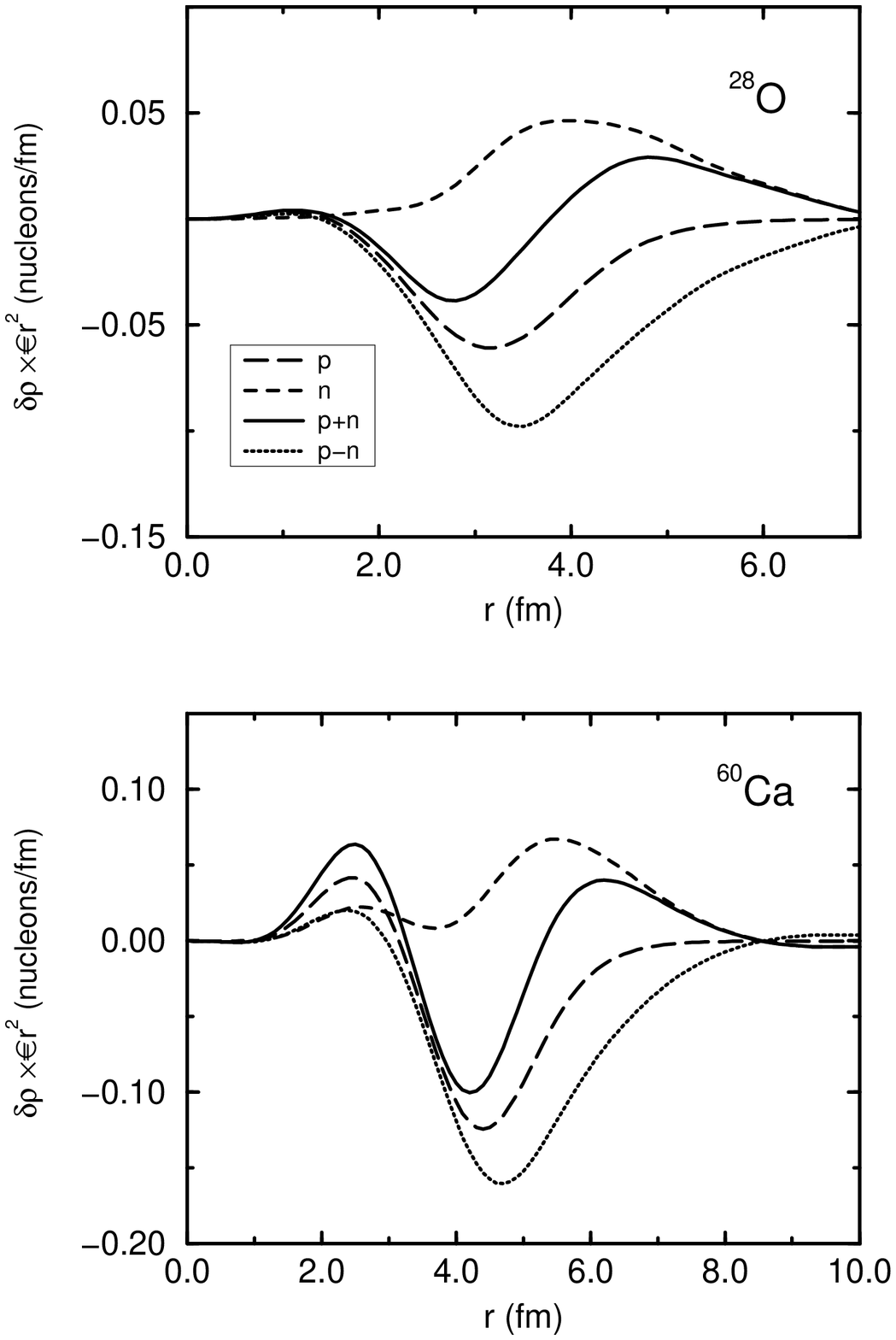}
\end{center}
\caption 
{Transition densities to selected RPA states.  Both isoscalar and
isovector densities are shown, together with the separate proton and
neutron contributions.  The states selected in $^{28}$O and $^{60}$Ca
have 18 and 17.4 MeV of excitation energy and carry 8 and 16 \% of the
EWSR, respectively. All densities are multiplied by $r^2$. }

\end{figure}

We can compare the microscopic transition densities with those
predicted by the macroscopic Goldhaber-Teller transition density

$$
\delta\rho_{GDR}^{isoscalar}(r)~=\delta\rho_{GDR}^p+\delta\rho_{GDR}^n=
~\alpha_1~\left[{2N \over A}~{d \rho_p \over dr}-{2Z 
\over A}~{d \rho_n \over 
dr}
\right]
$$
\noindent
where $\alpha_1$, given by
$$
\alpha_1^2~=~{\pi \hbar^2 \over 2m}~{A \over NZE_x}
$$ 
\noindent 
is the amplitude of the oscillation, derived from the
dipole EWSR.  $E_x$ is the excitation energy of the dipole state and
it is assumed that the state exhausts the full EWSR.  The feautures of
this isoscalar transition density can be better evidenced by expanding
it in the neutron excess parameter, according to Satchler~\cite{7}, in
the form
$$
\delta\rho_{GDR}^{isoscalar}(r)~\approx
~\alpha_1\gamma\left({N-Z \over A}\right)\left[{d 
\rho(r) \over dr}+
{R_0 \over 3}{d^2 \rho(r) \over dr^2}\right]
$$ 
\noindent 
where $R_0=(R_n+R_p)/2$ is the average radius of the
total nuclear density $\rho(r)$ and the parameter $\gamma$ is related
by $\gamma (N-Z)/A~=~3/2~(\Delta R/ R_0)$ to the measure of the
neutron skin $\Delta R~=~R_n-R_p$.  The isoscalar transition density
is therefore, to leading order, directly proportional to the neutron
skin.

\begin{figure}
\begin{center}
\includegraphics[bb= 60 60 550 719,angle=-90,scale=0.7]{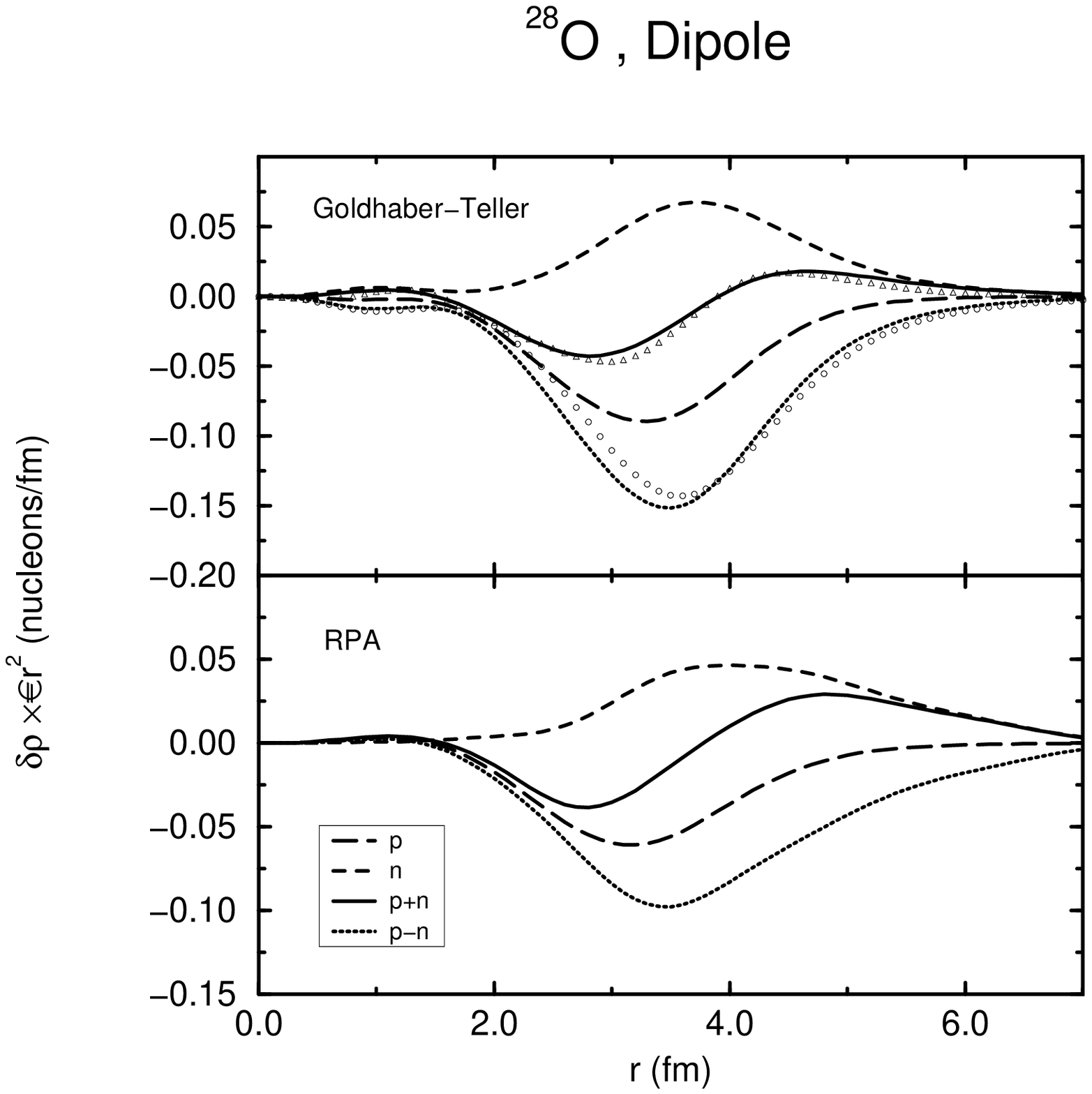}
\end{center}
\caption 
{Comparison of the microscopic proton and neutron transition densities
and their isoscalar and isovector combinations (lower part) with the
corresponding macroscopic expressions obtained within the
Goldhaber-Teller model (upper part).  The figure refers to the state
in $^{28}$O, shown in Fig.~5.  We also show in the upper figure as
curves with triangles and circles the approximated expression of
Satchler for the isoscalar and isovector transition densities,
respectively. In this approximate expressions, a value $\Delta R/R$ =
0.186 was used, taken from the HF density calculation. }

\end{figure}

The corresponding isovector transition density has a weaker dependence
on the neutron excess, to first order in $\Delta R/R_0$ and $(N-Z)/A$,
given by
$$
\delta\rho_{GDR}^{isovector}(r)~=\delta\rho_{GDR}^n-\delta\rho_{GDR}^p
$$
$$
=-~\alpha_1~\left[{2N \over A}~{d \rho_p \over dr}+{2Z 
\over A}~{d \rho_n \over dr}\right]
\approx
-~\alpha_1 ~{4NZ \over A^2}~{d \rho(r) \over dr}
$$

In Fig.~6, we show the comparison between the microscopic transition
densities and the macroscopic ones for the state in $^{28}$O, already
illustrated in Fig. 5.  The amplitude $\alpha_1$ of the macroscopic
model has been rescaled according to the proper percentage of the EWSR
exhausted by the state.  Both the ``exact'' Goldhaber-Teller
expressions and the approximated forms suggested by Satchler are used
(note that these latter practically coincide with the exact ones).
One can see from the comparison that the relevant features of the
microscopic RPA transition density are well reproduced by the
collective GT model.

One can observe from Figs. 5 and 6 the occurrence in the case of very
neutron-rich systems of a node in the isoscalar transition density at
the nuclear surface.  Furthermore, at large radii both the isoscalar
and isovector densities have similar radial dependence and magnitude.
This is a reflection of the fact that in this region it is only the
tail of the neutron density that contributes and thus both isoscalar
and isovector components contribute equally in this region.  The fact
that the isoscalar dipole transition density has different sign at
small and large radii has the consequence that the Coulomb-nuclear
interference will be destructive at small radii and constructive at
larger radii beyond the node.  This has been pointed out by several
authors~\cite{5,6,7} and this feature has been exploited as a specific
tool in order to obtain a measure of the neutron skin~\cite{12,13}.

\begin{figure}
\begin{center}
\includegraphics[bb= 110 110 550 719,angle=-90,scale=0.5]{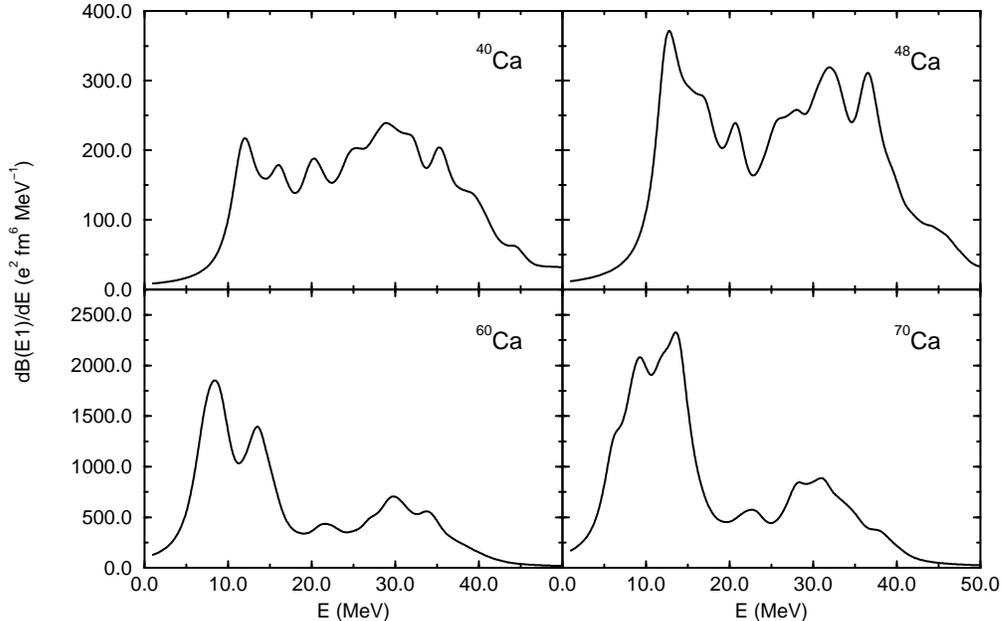}
\end{center}
\caption 
{Isoscalar dipole response to the RPA states obtained using the
operator $\sum_i z_ir_i^2$, for a sequence of Calcium isotopes, after
averaging the discrete spectra with a lorentzian with $\Gamma$=3
MeV. }

\end{figure}

\begin{figure}
\begin{center}
\includegraphics[bb= 60 60 550 719,angle=0,scale=0.9]{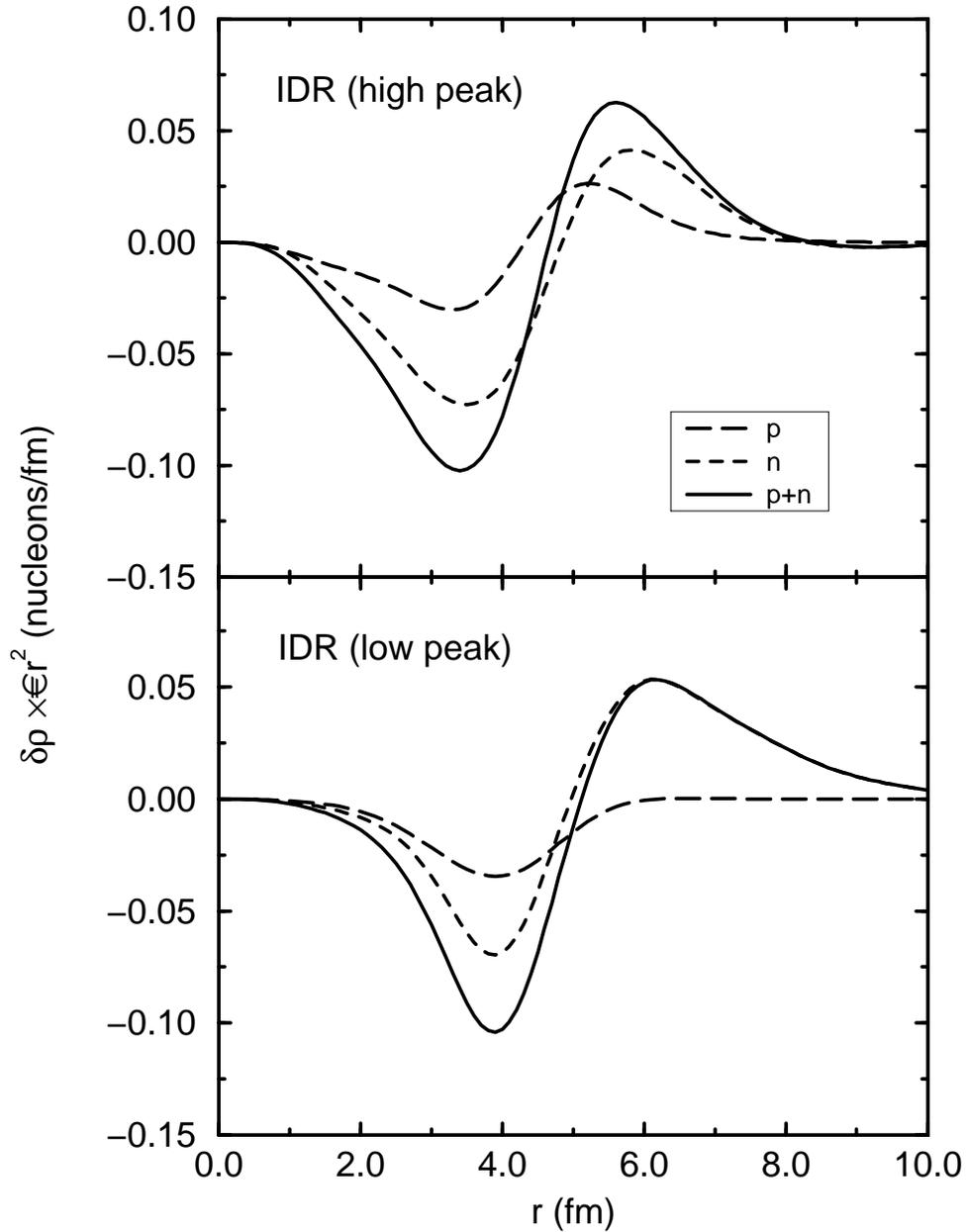}
\end{center}
\caption 
{Isoscalar transition densities to the selected states in $^{60}$Ca
collecting a large fraction of the isoscalar dipole strength.  The
state in the upper part (E=31.3 MeV) belongs to the high-energy peak
of the strength distribution, the one in the lower part (E=9.4 MeV) to
the low-energy peak.  The isoscalar transition densities are shown
together with the separate proton and neutron components. }

\end{figure}

Finally, we consider the possible compressional Isoscalar Dipole
Resonance (IDR) which is generated by the operator $\sum_i z_ir^2_i$.
This operator is the leading non-spurious term in the expansion of
$j_1(qr)Y_{10}(\hat {r})$ in the electromagnetic field.  The
excitation of this isoscalar giant dipole resonance via inelastic
scattering of $\alpha$ particles has been investigated in
ref.~\cite{14}, and the model has been developed and compared to
microscopic RPA calculations in ref.~\cite{15}, who considered systems
with different masses in the stability region.  In Fig.~7, we present
the evolution of the response for this operator in RPA with increasing
number of neutrons, considering the sequence of isotopes
$^{40,48,60,70}$Ca. It can be observed that there is an increased
strength in the response as one moves to neutron-rich nuclei,
approximately according to the "standard" $A^{7/3}$ scaling predicted
by the EWSR.  The neutron skin has in fact only a mild effect on the
total EWSR, which is given by~\cite{14,15,16}
$$
EWSR~=~{\hbar^2 A \over 8m\pi}\left[11 \langle r^4 
\rangle - {25 \over 3}
\langle r^2 \rangle ^2\right]
$$
\noindent
and which can be in leading order estimated to be
$$
EWSR~\approx~
{3 \hbar^2 r_0^4 \over 14 m \pi}~
A^{7/3}\left(1+2~{N-Z \over A}~
{\Delta R \over R}\right)
$$ \noindent 
So, even in the extreme case of $^{70}$Ca, the correction
over the $A^{7/3}$ scaling is of the order of few percent.  On the
other hand, aside from the EWSR, the large neutron excess seems to
have dramatic effects on the energy distribution of the strength.
While, in fact, along the stability valley the increased mass is just
shifting the collective peak to lower energies approximately according
to a $150 A^{-1/3}$ law~\cite{15}, in this case a splitting of the
strength in two main components occurs.  The lower one, which is
associated with the excitation of the neutrons of the skin, becomes
progressively more pronounced as one approaches the drip line.

This feature is better evidenced by the corresponding transition
densities.  We show in Fig.~8 the transition density for a state in
the low-energy peak and one in the high-energy peak. In the former
case the neutron contribution is completely dominant and the
transition density has a longer tail as expected for a contribution
involving weakly bound particles of the skin.  In the latter case the
neutron and proton components are comparably important, and they decay
faster in the tail and they are peaked at smaller distances, according
to the fact that they correspond to excitation of particles of the
core.

To summarize, we have studied within HF plus RPA with Skyrme
interaction the effects of increasing neutron excess on the isovector
Giant Dipole Resonance and on the compressional Isoscalar Dipole
Resonance. For the former we have found an increasing fragmentation of
the strength distribution when moving towards the drip line, together
with a shift to lower energies larger than the mass dependence
prediction based on collective models. For the latter, associated with
the isoscalar dipole operator $\sum_i z_ir^2_i$, a strong
concentration of the strength at low energy (around 10 MeV) has been
found for large neutron excess. This is interpreted as arising from
the excitation involving the neutrons of the skin.

\ack
One of us (F.C.) is grateful to the Departamento de FAMN 
of the University
of Sevilla for the warm hospitality during his stay made 
possible by
grant n$^{o}$ ERBFMBICT-961042 by the European Community 
within
the TMR program.

\end{document}